\centerline{\bf FRACTAL STRINGS AS THE BASIS OF CANTORIAN-FRACTAL}
\centerline{\bf SPACETIME AND THE FINE STRUCTURE CONSTANT}
\bigskip
\centerline{Carlos Castro}
\centerline {Center for Theoretical Studies of Physical Systems}
\centerline {Clark Atlanta University}
\centerline {Atlanta, Georgia, 30314}
\centerline {April, 2002}
\bigskip
\centerline {\bf Abstract}
\bigskip
Beginning with the most general fractal strings/sprays construction
recently expounded in the book by Lapidus and Frankenhuysen, it is shown
how the $complexified$ extension of El Naschie's Cantorian-Fractal
spacetime model belongs to a very special class of families of fractal
strings/sprays whose scaling ratios are given by suitable pinary (pinary,
$p$ prime) powers of the Golden Mean. We then proceed to show why the
logarithmic periodicity laws in Nature are direct physical consequences
of the complex dimensions associated with these fractal strings/sprays.
We proceed with a discussion on quasi-crystals with p-adic internal
symmetries, von Neumann's Continuous Geometry, the role of wild topology
in fractal strings/sprays, the Banach-Tarski paradox, tesselations of the
hyperbolic plane, quark confinement and the Mersenne-prime hierarchy of
bit-string physics in determining the fundamental physical constants in
Nature.

\bigskip

\centerline {\bf 1. Introduction: Fractal strings}

\bigskip

We will briefly summarize the basic ideas behind the book by [1] on
Fractal strings. A standard fractal string ${\cal L}$ is a bounded open
subset $\Omega$ of the real line $R$. It is well known to the experts
that such a set consists of countably many open intervals, the lengths of
which will be denoted by $l_1, l_2, l_3....l_j $. These are called the
lengths of the string. The sum $\sum l_j $ is finite and equals the
Lebesgue measure of $\Omega$. Despite the fact that one is dealing with
countably many intervals/lengths in the definition of $\Omega$ the
$boundary$ $\partial \Omega$ is not necessarily countably finite. The
$boundary$ of the Cantor string is the ternary Cantor set which is a
$non$ countable dust of points (it has the same cardinality as the real
line) despite having zero measure.

The geometric counting function of the lengths is defined
as: $$\zeta_{\cal L} (s) = \sum_{j = 1}^{\infty} (l_j)^s.
\eqno (1)$$ The $central$ notion of the book [1] is that
this function of $s$ has poles at complex numbers $s_n$,
the so-called complex dimensions of a fractal string.  The
spectrum of a fractal string consists of the sequence of
frequencies: $ f = k/l_j $ for $ k, j = 1, 2, 3...$. The
spectral zeta function of the fractal string $\cal L$ is
defined as:
$$ \zeta_\nu (s) = \sum_f f^{-s}. \eqno (2) $$
The geometry and the spectrum of $\cal L$ are connected by the following
relationship, valid for all values of $s$:
$$ \zeta_\nu (s) = \zeta_{\cal L} (s) \zeta_R (s). \eqno (3) $$
where the Riemann zeta function is:
$$\zeta_R (s) = \sum n^{-s}. \eqno (4) $$
it has a simple pole at $ s = 1 $ with residue $1$ and it admits an
analytic continuation to the whole complex plane.
The Cantor string is defined by constructing the sequence of lengths
using two scaling factors $ r_1 = r_2 = 1/3 $. One starts with the unit
interval and scales it by these two scaling factors yielding two segments
of length equal to $1/3$. One then repeats this process iteratively
yielding the Cantor string which is a self similar string consisting of
segments of lengths $ 3^{-n} $ with multiplicities of $ 1, 2, 4,...2^n
$ respectively. The geometric counting function of the Cantor string is
then given by the geometric series:
$$
\zeta_{CS} (s) = \sum 2^n 3^{- n s} = {1 \over 1 - 2\cdot
3^{-s}}. \eqno (5)
$$
The complex dimensions are given by the poles of this function, zeros of
the denominator:
$$ 2 3^{-s} = 1 = 1 e^{i 2 \pi n}. \eqno (6) $$
Taking logarithms on both sides of this equation yields:
$$ \ln 2 - s \ln 3 = 0 \pm i 2 \pi n \Rightarrow s =
{\ln2 \over \ln 3} \pm {i 2 \pi n \over \ln 3}. \eqno (7) $$
we immediately can recognize that the real part of the complex dimension
$ \ln 2/ \ln 3 $ coincides precisely with the fractal dimension of the
ternary Cantor set. The $boundary$ of the fractal string $ \cal L $ is
precisely the uncountable Cantor fractal dust of points.
There are many types of fractal strings discussed in the book [1].
In particular, the Fibonacci fractal string which is constructed using
the scaling ratios $ r_ 1 = 1/2 $ and $ r_2 = 1/4 $. The sequence of
lengths are $ 1, 1/2, 1/4. 1/8,.....1/2^n,...$ with respective
multiplicities given by the Fibonacci numbers $ 1,1, 2, 3,....F_{n+1
}...$ which obey $ F_{n+1} = F_n + F_{n-1}$ and whose ratio in the large
$n$ limit is given by the Golden ratio:
$$ \lim_{n \rightarrow \infty} ~{F_{n+1} \over F_n} = \tau = 1+\phi = {
1 + \sqrt 5 \over 2} = 1.618...\eqno (8) $$
The geometric counting function of the Fibonacci string is:
$$ \zeta_{Fib} (s) = {1 \over 1 - 2^{-s} - 4^{-s}}. \eqno (9) $$
and the poles are found by solving $ (2^{-s})^2 + 2^{-s} = 1 $ which
gives the complex dimensions:
$$ {\ln \tau\over \ln 2} \pm {i 2 \pi n \over \ln 2} $$
$$ - {\ln \tau \over \ln 2} \pm {i 2 \pi (n + 1/2) \over \ln 2}.
\eqno (10) $$
Notice that $negative$ dimensions are very natural ingredients of the
complex dimensions of fractal strings. Negative dimensions fit the
negative-entropy proposal of Michael Conrad [3] to understand living
systems.
Both the Cantor and Fibonacci strings belong to the class of lattice
strings, meaning that they have oscillations in the geometry of the order
of $ D = Real (s) $. We can hear the dimension of fractal strings [1]
. We can also hear if a string is Minkowski measurable if, and only if,
the $\zeta (s) $ has $no$ zeros on the line $ Real (s) = D $. Cantor
strings, lattice strings, are not Minkowski measurable.
A class of nonlattice strings is given, for example, for scaling ratios
given by $ r_1 = 1/2 $ and by $ r_2 = 1/3 $; i.e the scaling ratios are
not given by powers of a fixed $ r < 1 $. Another example is given by the
so-called Golden string in [1] whose scaling rations are $ r_ 1 = 1/2 $
and $ r_2 = (1/2)^{1 +\phi} $. In particular, the complex dimensions of
these strings are given by solving a $transcendental$ equation which
yields a $nonlattice$ structure of points in the complex planes. For this
reason these strings are called non-lattice, despite being self-similar.
The complex dimensions of the Golden string is given by an almost
periodic structure in the complex plane. All complex dimensions of a self
similar string with scaling ratios $ r_1, r_2,...r_N$ lie to the left
of, or on the line
$ Real (s) = D$ [1].
In order not to confuse the reader with the nomenclature used in [1] we
emphasize that our construction of Cantorian-Fractal spacetime [2] is
based entirely in a particular class of fractal strings, and higher
dimensional fractal sprays or branes, based on suitable binary powers and
pinary powers of the Golden mean. By pinary we mean powers of a prime
number $p$:
$$ 2^{- \phi^j}. ~~~ p^{- \phi^j}. ~~~ j = \pm 1, \pm 2, \pm 3....
\eqno (11) $$
In section {\bf 3} we will show that complex dimension is
not just a mathematical artifact but that is deeply related
to the log-periodic laws in Nature discussed amply by
Nottale et al [4,5] in their theories of the Fractal Tree
of life and Fractal Evolution and by many others in
particle physics in the renormalization group context
[6,7]. We believe that quantum gravitational phenomena
should involve $interference$ of complex dimensions. With
this preamble of some of the basic ideas on fractal strings
of the book [1] we turn attention to the construction of El
Naschie's Cantorian-Fractal spacetime and its
complexification. Before doing so, we must discuss the
notion of fractal sprays (branes) which are the higher
dimensional analogs of strings.
\bigskip
\centerline {\bf 2. Fractal Branes/Sprays}
\bigskip
In this section we are going to generalize the construction of fractal
strings to the $p$-branes case. In particular we will be studying a
subclass of fractal branes called $sprays$ in the literature [1]. Having
done so we will show how the Cantorian-Fractal spacetime model [2] is a
very special representative of these fractal sprays models given by [1]
and which are based on scaling ratios given by powers of the golden mean
$ \phi^j $, where $j = \pm 1, \pm 2, \pm 3,.....$.
A self-similar fractal spray [1] $ \Omega$ with basic shape
$B$, scaled by a self-similar string $ \cal L$, whose dimension can be
$greater ~than~ one $, is given by a collection of sets $\Omega_j$ which
are congruent to $ l_j B $, the homothetic of $\Omega$ by the ratio $
l_j < 1 $ for each value of $ j $. Hence, for instance, the fractal spray
of the Cantor string on the unit square $ B $, see figure, is obtained by
having one open unit square; two open squares of length-size $ 1/3 $,
four open squares of length-size $ 1/9 $...and so forth.
The spectral zeta function of the Dirichlet Laplacian on the square is:
$$ \zeta_B (s) = \sum_{n_1, n_2} (n^2_1 + n^2_2)^{-s/2}. \eqno (12
) $$
and the spectral zeta function of the spray associated with the
Cantor-string is defined by:
$$ \zeta_\nu (s) = \zeta_{CS} (s) \zeta_B (s) \eqno (13) $$
where the geometric length counting function of the Cantor string was
given in the previous section:
$$ \zeta_{CS} (s) = \sum_j (l_j)^s = \sum 2^n 3^{- n s} =
{1 \over 1 - 2 3^{-s}}. \eqno (14) $$
The $ \zeta_B(s) $ has poles at $ s = 2 $ and the $\zeta_{CS} (s) $
has poles at: $ s = (\ln 2/\ln 3) \pm i 2 \pi n / \ln 3 $.
The Sierpinski drum is also a fractal spray associated with the unit area
triangle $ {\cal T} $ and is obtained by scaling the middle triangle with
the scaling ratios: $ r_1 = r_2 = r_3 = 1/2 $. Notice now that $ \sum
r_j = 3/2 > 1 $ and for this reason the dimension of this fractal spray
will be greater than unity. There is one triangle of unit area, $3$
triangles of $ 1/4 $ area, $ 9 $ triangles of $ 1/16$ area, and so forth.
The geometric $length$ counting function associated with the scaling
lengths $ r_1 = r_2 = r_3 = 1/2 $ is:
$$
\zeta_{\cal L} (s) = \sum (l_j)^s = \sum 3^n 2^{- n s} =
{1 \over 1 - 3\cdot 2^{-s}}. \eqno (15)
$$
and has poles at:
$$ s = {\ln 3 \over \ln 2} \pm {i 2 \pi n \over \ln 2}. \eqno (16
) $$
Notice that the real part of $ s $ is given by $ D = \ln3/\ln 2 > 1 $
which
is precisely the fractal dimension of the Sierpinski's gasket.
The spectral zeta function of the Dirichlet Laplacian on the unit
triangle is:
$$ \zeta_{\cal T} (s) = \sum_{m, n} (m^2 + mn + n^2)^{-s/2}. \eqno
(17) $$
and it has poles at $ s = 2 $ with residue $(\pi/3\sqrt 3) $ and at $ s
= 1 $ with residue $ -3/4 $. The spectral zeta of the Sierpinski drum is
finally given by:
$$ \zeta_\nu (s) = \zeta_{\cal T} (s) {1 \over 1 - 3 2^{-s}}.
\eqno (18) $$
Notice that $no$ complex dimension of $ {\cal L}$ (the string whose
scaling ratios $ r_1 = r_2 = r_3 = 1/2$ were used to generate the
Sierpinski drum whose real dimension is given by $ D = \ln3/\ln2 $)
$ coincides$ with the poles of $ \zeta_{\cal T} (s) $, given by $ s = 2
$ and $ s = 1$. For this reason $ D $ is also the dimension of the
$boundary$ of the fractal spray $\Omega $ associated with the string $
{\cal L} $. Hence we have that:

$$ d - 1 \leq D = dim~ \partial \Omega \leq d \Rightarrow 1 \leq {\ln3
\over \ln 2} \leq 2. \eqno (19) $$ where $ d = 2 $ is the dimension of
the ambient embedding space $ R^2 $ used in the construction of the
fractal spray $\Omega $.

\bigskip

\centerline {\bf 3. Cantorian-Fractal spacetime as a very special class of
Fractal strings/sprays}

\bigskip

Having outlined how to construct a fractal spray in the ambient space $
R^2 $ will allow us to show why Cantorian-Fractal spacetime

$ {\cal E}^{(\infty)} $ [2] comprised of an infinite hierarchy of
sets
$ {\cal E}^{(j)} $ of dimension:
$$ (1 + \phi)^{j -1}. ~~~ j = 0, \pm 1, \pm 2, \pm 3.....\pm \infty
. \eqno (20) $$
is a special class of fractal strings/sprays [1] whose scaling ratios
are suitable binary powers of the Golden mean
$ 2^{- \phi^{j - 1}} $. We shall call these fractal sprays the Golden
strings/sprays and must $not$ be confused with the Golden strings
discussed by the authors [1]. The latter are $nonlattice$ self similar
strings whereas the former are lattice self similar ones.
Let us consider the Golden fractal spray $ \Omega $ obtained by scaling
an open square $B$ of unit area by the scaling lengths:
$$ r_1 = r_2 = 2^{- \phi}. \eqno (21) $$
Thus $ \Omega $ is a bounded open subset of $ R^2 $ consisting of one open
square of unit area, $2$ open squares of length-size
$ 2^{- \phi} $ (area is $ 2^{- 2 \phi} $), $ 4 $ open squares of
length size $ 2^{- 2 \phi} $ (area is $ 2^{- 4 \phi} $), etc....
The geometric length counting function associated with the scaling:
$ r_1 = r_2 = 2^{- \phi} $ is:
$$ \zeta_{\cal L} (s) = \sum (l_j)^s = \sum 2^n 2^{- n \phi s} =
{1 \over 1 - 2 2^{- \phi s}}. \eqno (22) $$
and it has poles when the denominator vanishes:
$$ 1 = 1 e^{i 2 \pi n} = 2 2^{- \phi s} \Rightarrow 0 \pm i 2 \pi n =
\ln 2 - \phi s \ln 2 \Rightarrow $$
$$ s = (1 + \phi) \pm {i 2 \pi n (1 + \phi) \over \ln 2} \eqno (23)
$$
where we have used the defining relation of the Golden Mean:
$$ {1 \over \phi} = 1 + \phi \Rightarrow 1 = \phi + \phi^2 \Rightarrow
\phi = {\sqrt 5 - 1 \over 2} = 0.618... \eqno (24) $$
It is not difficult to prove that the Golden mean generates a ring
structure $ Z [\tau] $, where $ \tau = 1 + \phi $:
$$ (1 + \phi)^n = F_{n+1} + \phi F_n. ~~~
\phi^n = (- 1)^n F_{n-1} + (-1)^{n+1} F_{n} \phi. \eqno (25) $$
where $ F_n$ are the Fibonacci numbers $ 1,1, 2, 3, 5, 8, 13, 21...$
obeying the recursive relation: $ F_{n+1} = F_n + F_{n -1} $ and the
limit of $ F_{n+1}/F_n = 1 + \phi $ when $ n \rightarrow \infty$.
Ring structures are essential ingredients in von Neumann's formulation of
Continuous Geometry [8].
The spectral Dirichlet Laplacian of the unit square is:
$$ \zeta_B (s) = \sum (n^2_1 + n^2_2)^{- s/2} \eqno (26) $$
and it has poles at $ s = 2 $.
Hence the spectral zeta function of this fractal Golden spray is:
$$\zeta_\nu (s) = {1 \over 1 - 2 2^{- \phi s}} \zeta_B (s).
\eqno (27) $$
Since the poles of $ \zeta_B(s) $ do $ not$ coincide with
the zeros of the denominator $ 1 - 2 2^{-\phi s}= 0 $ we have that the
dimension of the Golden spray is precisely equal to the dimension of the
$boundary$ of the open bounded domain $\Omega$: $\partial \Omega$. Hence
we have that:
$$ d - 1 \leq (1+\phi) \equiv dim~\partial \Omega \leq d \Rightarrow 1
\leq 1+ \phi \leq 2. \eqno (28) $$
This procedure can be generalize to higher-dimensions and in this fashion
we will be able to construct a physical model of Cantorian-Fractal
spacetime [2] from the fractal sprays associated with the basic domains
given by the unit size hypercubes of increasing dimensionality.
Since $ 1+\phi < 2 $ we were able to construct the set
$ {\cal E}^{(2)}$ based on scaling of two-dim square domains. Because
$ (1+\phi)^2 = 2 + \phi < 3 $ we can construct the set
$ {\cal E}^{(3)}$ by a similar procedure of scaling the $ 3$-dim unit
cube by factors of $ r_1 = r_2 = 2^{- \phi^2} $.
The geometric length counting function in this case is:
$$ \zeta_{\cal L} (s) = \sum 2^n 2^{- n \phi^2 s} =
{1 \over 1 - 2 2^{- \phi^2 s}}. \eqno (29) $$
and has poles at the zeros of the denominator:
$$ 1 = 1 e^{i 2 \pi n} = 2 2^{- \phi^2 s} \Rightarrow
0 \pm i 2 \pi n = \ln 2 - \phi^2 s \ln2 \Rightarrow $$
$$ s = (1 + \phi)^2 \pm {i 2 \pi n (1 + \phi)^2 \over \ln 2}.
\eqno (30) $$
Hence we can see that the real part of the complex dimension coincides
with the corresponding dimension of the set:
$$ (1 + \phi)^2 = 2 + \phi = dim~ {\cal E}^{(3)}$$
The spectral zeta function of the Dirichlet Laplacian on
the unit cube is:
$$ \zeta_B (s) = \sum (n^2_1 + n^2_2 + n^2_3)^{-s/2}. \eqno (31)
$$
and it has poles at $ s = 3 $. The dimension of this Golden spray
coincides with the dimension of the $boundary$ of the $3$-dimensional
open domain $\Omega$ living in the ambient space $ R^3 $:
$$ d - 1 \leq (1+\phi)^2 \equiv dim~\partial \Omega \leq d \Rightarrow 2
\leq (1+ \phi)^2 = 2 + \phi \leq 3. \eqno (34) $$
This construction generalizes to the full space
$ {\cal E}^{(\infty)} $ with the provision that one constructs the
Golden sprays from suitable hypercubes of unit hypervolumes and of
$enough$ dimensionality $N$ to obey:
$$ N > (1 + \phi)^n = F_{n+1} + \phi F_n > N -1. \eqno (35) $$
For example, to construct the set $ {\cal E}^{(4)}$ whose dimension is
$ (1 + \phi)^ 3 = 4 + \phi^3 = 3 + 2 \phi > 4 $ requires starting with
a unit size hypercube in $ R^5 $, instead of $ R^4 $ !.
To construct the set $ {\cal E}^{(5)}$ whose dimension is
$ (1 + \phi)^4 = 5 + 3 \phi > 6 $ requires starting with a unit size
hypercube in $ R^7 $. And in general to construct the set
${\cal E}^{(j)} $ whose dimension is $ (1 + \phi)^{j-1} $ requires to
find the smallest integers $ N (j) $ such that:
$$ N (j) ~> ~(1 + \phi)^{j-1} = F_j + \phi F_{j-1} ~> ~ N(j) - 1.
\eqno (36) $$
and to begin the construction of the Golden spray by starting with a unit
size hypercube in $ R^{N (j)} $ whose scaling ratios are:
$$ r_1 = r_2 = 2^{- \phi^{j-1}}.~~~ j =2,3, 4, 5,.... \eqno (37) $$
The poles of the corresponding geometric counting length:
$$\zeta_{\cal L} (s) = \sum 2^n 2^{- n \phi^{j-1} s} =
{1 \over 1 - 2 2^{- \phi^{j-1} s}}. \eqno (38) $$
are obtained as usual, from the zeros of the denominator:
$$ 1 = 1 e^{i 2 \pi n} = 2 2^{- \phi^{j-1} s} \Rightarrow
0 \pm i 2 \pi n = \ln 2 - \phi^{j-1} s \ln2 \Rightarrow $$
$$s = (1 + \phi)^{j-1} \pm {i 2 \pi n (1 + \phi)^{j-1} \over \ln 2}.
\eqno (39) $$
Hence, the dimension of the fractal $boundary$ of the domain $ \Omega $
living in the ambient space $ R^{N (j)} $ coincides precisely with
the dimension of the set $ {\cal E}^{(j)} $:
$$ N(j) - 1 \leq (1+ \phi)^{j -1} = dim~ \partial \Omega \leq N(j).
\eqno (40) $$
This procedure also works for sets whose dimensionality is
less than one:
$$ 0 \leq D \leq 1. \eqno (41) $$
By starting with the interval $ (0, 1) $ one may construct a hierarchy
of fractal strings of dimensionality:
$$ 0 \leq \phi^{j - 1} \leq 1. \eqno (42) $$
by using the scaling ratios $r_1 = r_2 = 2^{- (1+\phi)^{j-1}} $.
The corresponding geometric counting lengths of this hierarchy of fractal
strings are:
$$ \zeta_{\cal L} = \sum 2^n 2^{- n (1+\phi)^{j-1} s} =
{1 \over 1 - 2 2^{- (1+\phi)^{j-1} s}}. \eqno (43) $$
whose poles are located at:
$$ s = \phi^{j -1} \pm {i 2 \pi n \phi^{j-1} \over \ln 2}. \eqno (44)
$$
The dimensions of this hierarchy of fractal strings is $ 0 \leq D \leq 1
$ agree with the values of the dimension function of von Neumann's
Continuous Geometry [8].
Notice that the normal set $ {\cal E}^{(1)} $ has dimension equal to
$ D = (1 + \phi)^{1 - 1} = 1 $. This set corresponds to the interval
$ (0,1)$. The scaling ratios in this limiting case are $ r_1 = r_2 =
1/2 $ which entails that if we scale the unit interval by $r_1 = r_2 =
1/2$ and join-in the $two$ segments of length $1/2$ we will get back the
unit interval. Repeating this procedure ad infinitum one we will always
end back with the unit interval. This normal set is not fractal.
Concluding, we have shown how Cantorian-Fractal spacetime belongs to a
very special class of fractal strings/sprays whose scaling ratios are
given by suitable binary powers of the Golden Mean. This procedure
automatically furnishes the topology of ${\cal E}^{(\infty)} $
displayed by an infinite collection of sets as show in figure, for the
set $ {\cal E}^{(2)}$.
We could have built the fractals sprays differently, by
scaling given by:
$$ p^{- \phi^{j-1}}. ~~~ p = prime. \eqno (45) $$.
For example, taking $ p = 3 $ our basic domain would have been a triangle.
Then we scale this triangle by the scaling functions $ r_1 = r_2 = r_3 =
3^{- \phi} $ to generate $ 3 $ congruent triangles of areas $ 3^{- 2
\phi}$ smaller than the original and that are attached symmetrically to
the $3$ sides of the unit triangle. Repeating this process ad infinitum
will yield $9, 27,...3^n..$ scaled-down versions of the unit triangle and
allows to construct a model of the set $ {\cal E}^{(2)}$ as well. The
geometric counting function will be:
$$ \zeta_{\cal T} (s) = \sum p^{n} p^{- n \phi^{j -1} s} =
{1 \over 1 - p p^{- \phi^{j-1} s}}. \eqno (46) $$
whose poles are at:
$$ s = (1 + \phi)^{j -1} \pm {i 2 \pi n (1 + \phi)^{j -1} \over \ln
p}. \eqno (47) $$
The real part of the dimension coincides again with the dimension of the
set $ {\cal E}^{(j)} = (1 + \phi)^{j -1} $. Only the periods of
the imaginary components will vary accordingly to different values of $ p
= 2, 3, 5, 7,...$. It is essential to use scaling given by powers of
primes: $pinary$ powers.
These construction of these fractal sprays by scaling of the unit size
domain and their subsequent iterated attachings resembles nothing but a
$crystal$ growth process; i.e and assembly of hypersurfaces. This
assembly of surfaces occurs in the theory of $capped~gropes$ and link
homotopy associated with four-manifolds and belongs to a branch of
mathematics called Geometric Topology. The number of surfaces that are
attached grow like $ 2^h - 1 $, where $ h$ is the grope height. One
could envision these fractal sprays as nontrivial embedding of circles
and (higher-dimensional) spheres into the ambient spaces
$ R^n$. The most famous examples of nontrivial embedding are the
Antoine's necklace (a wild knot) and Alexander's Horned Spheres. Their
exteriors in $R^3 $ are $not$ simply connected and hence are not
homeomorphic to the standard embedding of a circle and a sphere. For this
reason, these objects belong to what is called wild topology. For recent
work on this topic see [9].

In the last section we will discuss further the $p$-adic Topology and the
$p$-adic internal symmetries of quasi-crystals and their relation to the
Fibonacci modules, the Dirichelt zeta, function, the ring of $Z[\tau] $
[10]. It is this subject that will bring us to bit-strings physics, the
Mersenne prime hierarchy, Set Theory, the Banch-Tarski paradox,
tesselations of hyperbolic planes...and other topics in relation to quark
confinement and the values of the fundamental constants in Nature.

\bigskip
\centerline {\bf 4. Log-periodicity as the physical basis of Complex
Dimensions}

\bigskip

In the past years a lot of activity has been concentrated on the
log-periodic laws associated with the fractal structures of evolutionary
trees. In particular, the time sequences of major evolutionary leaps at
various time scales [5]. Models of this type have been observed in
economical crisis patterns in Western pre-Columbian civilizations. The
physical model underlying the appearance of such laws is that of critical
phenomena [6,7]. The Renormalization Group approach predicts both power
laws and logarithmic-periodic corrections.

Nottale et al [5] have considered the simplest Galilean-like
renormalization group like equations for the variations of a
non-differentiable fractal function $ L (x, \epsilon) $ than depends
on $ x$ and the resolutions $ \epsilon $ with respect to logarithmic
scaling:
$$ {\partial L (x, \epsilon) \over \partial \ln \epsilon} = \beta (L)
. \eqno (46) $$
Assuming a ``beta'' function proportional to $ L $:
$$ \beta (L) = \delta L \Rightarrow
L (x, \epsilon) =
L_o (x) ({\epsilon_o \over \epsilon})^\delta. \eqno (48) $$
where $ \epsilon_o = \lambda $ is the transition ``scale'' from a
fractal
to non-fractal behavior of the physical system under consideration. The
scaling exponent $ \delta = D - D_T $, where $ D_T $ is the topological
dimension and $ D $ is the fractal dimension. A space-filling curve (a
Peano curve) has $ D_T = 1 $ but $ D = 2 > D_T$.
If the fractal dimension $ D$ is complex valued, $ d_x + i d_y $, then
one can see that the imaginary component will be responsible for an
oscillatory behavior which yields the logarithmic-periodic character to
the function:
$$ L = L_o (x) ({\epsilon_o \over \epsilon})^{i d_y} = L_o (x) \exp
[i d_y \ln (\epsilon_o/\epsilon)] \Rightarrow $$
$$ L = L_o (x) \cos [d_y \ln ({\epsilon_o \over \epsilon})
+ \alpha]. ~~~ \alpha = phase. \eqno (49) $$
For example, setting the phase factor to zero and using the complex
dimensions of the class of fractal strings related to $ {\cal E}^{
(\infty)}$, we have the following hierarchy of the imaginary components
of the complex dimensions:
$$ d^{(n)}_y = {2 \pi n \over \ln 2} (\phi)^{j - 1}. \eqno (50) $$
Following the Fractal Tree of life description of [5] one has the
following identification:
$$ \epsilon_0 = T_0 - T_c. ~~~ \epsilon = T - T_c. \eqno (51)$$
where $ T_c$ is the critical time marking the end of an evolutionary
process which began at $ T_0 $. And $T$ is the time variable. One can see
that the function $ L$ has peaks at discrete values of time $ T^{(j)
}_n$
that accelerate toward the critical dates according to a log-periodic
law:
$$ {T^{(j)} _n - T_c \over T_0 - T_c} = 2^{- n (1+ \phi)^{j -1}} <
1. \eqno (52) $$ where the label $ j $ is associated with the
particular string or branch $ {\cal L}_j $, generated by the scaling
ratios $ 2^{- (1 + \phi)^{j -1}} $, and the label $ n $ represents
the level height within each particular string/branch. The arguments of
the cosine function will then be:
$$ [{2 \pi (\phi)^{j -1} \over \ln 2}] ~ \ln ~
2^{n (1+ \phi)^{j -1}} = (1 + \phi)^{j -1} \phi^{j -1} 2 \pi n =
2 \pi n.~~~~ \phi (1 + \phi) = 1. \eqno (53) $$
and hence, we will have peaks spaced at log-periodic intervals. We can
then model the fractal evolutionary tree process of Nottale et al [5]
in terms of fractal strings/sprays; i.e. in terms of a sequence of
temporal intervals of lengths $ T^{(j)}_n - T_c$ ! associated with a
given branch $ {\cal L}_j $. Hence, the fractal tree of life envision by
[5] fits very naturally within the context of fractal strings and their
complex dimensions described in the book
[1].

If Physics is supposed to explain all of natural phenomena then it must
be able to explain Biology and Evolution in Nature. The fractal paradigm
described here on the basis of fractal strings/sprays will be a nice
starting point. Since we live in four spacetime dimensions it would not
be surprising that $ d = 4$ will be a relevant scaling exponent.

\bigskip
\centerline {\bf 5. Further Topics: p-adic numbers, Quark Confinement
and the Fine Structure Constant}
\bigskip
In this last section we will discuss further the $p$-adic Topology and
the $p$-adic internal symmetries of quasi-crystals and their relation to
the Fibonacci modules, the Dirichelt zeta function and the ring of $Z[
\tau]$ [10] before embarking into other topics.

Quasi-periodic point sets have been interpreted by many authors as
quasi-crystals with $p$-adic internal symmetries and $p$-adic Topology
[10]. One can generalize the fractal strings/sprays construction of [1]
to spaces with $p$-adic Topology instead of Euclidean one. These will
be then called $p$-adic Fractal Strings and Sprays. Ordinary $p$-adic
strings have been studied extensively in the Literature [11]. The
reason why $p$-adic fractal strings may be very relevant at Planck scales
is due to the Non-Archimedean geometry: one cannot naively add/subtract
lengths as we ordinary do in ordinary Euclidean geometry, since the
Planck scale is taken in Extended Scale Relativity [12] to be the minimum
length. One requires to use an ultrametric $p$-adic distance which
reinforces the minimum length principle. In fact we have shown that based
on the ideas of C-spaces, Clifford manifolds, one can reproduce the
current results of kappa-deformed Poincare theories of gravity [12].

A large number of quasi-periodic tilings (tesselations of the plane,)
display these $p$-adic properties. A class of $ Z$-modules, the Fibonacci
modules, with both crystallographic structures and scaling invariance for
the ring $ Z [\tau] $, the ring of integers in the quadratic field $ Q
[\tau] $, where $ \tau = 1 + \phi $, have been studied by [10]. The
number of self-similar submodules was encapsulated by the Dirichlet zeta
series associated with the Fibonacci module $ M$:
$$ \zeta_{M} (s) = \sum {a (m) \over m^s} = \sum_{\tilde M}
{1 \over [M:{\tilde M}]^s}. \eqno (54) $$
where $ {\tilde M} $ runs over the self-similar submodules of $M$ and the
quantity defined as $[M: {\tilde M}]$ is the $index$ factor, that
may or may not bear any relation to Jones theory of subfactors in Knot
theory. One of the most interesting quasi-crystals are those related to $
4$ dimensions, like the Elser-Sloane quasi-crystals associated with the
Hurwitz ring of integral quaternions or the icosian ring [10]. It has
been know for some time that Penrose quasi-periodic tiling of the plane
is one of the most simplest examples of Noncommutative Geometry whose
$K$-theory groups are in fact related to the ring $ Z[\tau] $,
connected to the Golden Mean.
Using these ideas Selvam-Fadnavis [13] gave an estimate of the inverse
fine structure constant:
$$ 20 (1 + \phi)^4 = 100 + 60 \phi = 137. 082....\eqno (55) $$
based on Penrose quasi-crystal with a five-fold symmetry and a complex
growth-process involving the logarithmic spiral with the Golden Mean
winding number. There are $ 5 $ successive growth-steps in this process,
involving a clockwise and counter-clockwise motion giving a net number of
$ 5. 2 = 10$. The variance of each step is $ 2 \sigma = 2 (1 + \phi)^4$.
Hence the total variance is given by eq-(55) and was related to the
inverse fine structure constant. El Naschie [14] has obtained the same
value as [13] based on a transfinite heterotic string theory formalism,
interpreting the number $ 10 $ as the dimension of the superstring.

As suggestive as these proposals are, we are more inclined to adopt the
bit-string physics and combinatorial hierarchy approaches of the last
$30$ years by P. Noyes and collaborators [15]. The reason is based once
again in the notion of $p$-adic numbers, $p$-adic Topology,
quasi-crystals, lattices and the Mersenne prime hierarchy. A pure set and
lattice construction of $all$ the coupling constants in Nature using
discrete Clifford algebras was given by T. Smith [16]. A Mersenne prime
hierarchy was furnished by Pitkanen in his formulation of Topological
Geometrodynamics [17].
Based on the ideas of $p$-adic numbers and transfinite $M$ theory we have
shown that the sum of suitable powers of the Golden Mean 18]:
$$ 1 + (1 + \phi)^{2} + (1 + \phi)^{2^2} + (1 + \phi)^{2^3} +
(1 + \phi)^3 + (1 + \phi)^{3^2} = $$
$$ 100 + 61 \phi = 137.7....\eqno (56) $$
Our proposal was inspired from the $p$-adic decomposition of $ 137 $,
for $ p = 2 $ and has been reinforced lately by the bit-string physics
work:
$$ 137 = 3 + 7 + 127 = (2^2 - 1) + (2^3 - 1) + (2^7 - 1) $$ which
admits the $2$-adic (binary) expansion:
$$ 137 = 2^0 + 2^3 + 2^ 7 = 1 + 8 + 128. \eqno (57) $$
Notice that the sum of $ (3, 7, 127) $ is indeed $ 137$. The sum of $ 3
+ 7 = 10 $ (any connection to the dimension of superstrings?) and that
these numbers are nothing but Mersenne primes:
$$ M_p = 2^p - 1, ~~~ p = prime. \eqno (58) $$
A table of well known Mersenne primes is [19]:
$$ p = 2, 3, 5, 7, 13, 17, 19, 31, 61, 89, 107, 127, 521.... \eqno (59
) $$
The Mersenne primes $ 3, 7, 127 $ are very special because they belong as
well to the table of $doubly$ iterated Mersenne primes:
$$ M_2 = 2^2 - 1 = 3 \Rightarrow MM_2 = 2^3 - 1 = 7 \Rightarrow M_7 =
2^7 - 1 = 127 \Rightarrow M M_7 = 2^{127} - 1 = prime. \eqno (60) $$
It is unknown if the number:
$$ 2^{2^{127} - 1} - 1 $$
is a prime or not. The number $ 2^{127} - 1 $ is a $prime$ and is of the
order of $ 10^{38} $. These numbers obtained by powers of multiple
exponents are called googol-plexus. Bit-string results suggest that these
$four$ doubly iterated numbers is related to the the fact that there are
$four$ forces in nature. Speculations that the four division algebras of
octonions, quaternions, complex and reals may be connected to the four
forces have been invoked by some authors. We don't know if this is true or
not, all we can say is that is very suggestive.

The $p$-adic norm of $137$ for $ p = 2 $ is:
$$ || 137 ||_2 = {1 \over 2^0} = 1. \eqno (61) $$
the physical interpretation of this unit $2$-adic norm for $137$ can be
attributed to the bit-string physics result that $1$ Coulomb event
requires $ 137$ bits [15]:
$$ || 137 ||_2 = {1 \over 2^0} = 1 \Leftrightarrow 1~ Coulomb ~event =
137~ bits. \eqno (62) $$
The number $ 20 (1 + \phi)^4 = 100 + 60 \phi = 137.082 [13, 14] $
does $not$ admit a $p$-adic like expansion like the one given in eq- (56
). The interpretation of eq-(56) as sums of powers of the Golden Mean
admits also a fractal string/spray interpretation:
Each term in the expansion of eq- (56) corresponds precisely to the
fractal dimension of a fractal Golden string/spray. Hence, eq-(56) is
just:
$$ 100 + 61 \phi = \sum dim ~ {\cal E}^{(n)}. \eqno (63) $$
for those sets whose values of $n$ are given by:
$$ n = 1, 3, 4, 5, 9, 10 $$ respectively.
Therefore, the number proposed by us [18]
$ 100 + 61 \phi $ is just the sum of the fractal dimensions of a family
of fractal sprays constructed by using scaling ratios given by binary
powers of the Golden Mean. Our number is irrational and lies between
$138$ and $137$. Let us summarize the bit-string physics results.
The Mersenne prime plus the bit-string combinatorial factors give an
inverse fine structure value of:
$$ \alpha^{-1} = 137 [1 - {1 \over 30 \times 127}]^{-1} =
137.0359674. \eqno (64) $$
in astonishing agreement compared with the experimental results:
$ 137.0359895 $. Recent Astrophysical observations seem to suggest that
the fine structure is varying over cosmological times.
A variable fine structure constant based on $C$-spaces has been given by
us [12].

Notice that bit-string results involve a $finite$ number of concatenation
or string bits. The numbers obtained from bit-string physics are
$rational$ ones, whereas the fractal strings generate $irrational$
numbers due to their $transfinite$ nature. Irrational Conformal Field
Theory has barely been studied in physics. Deep relations to number
theory, conformal field theory and the Monster group exist as pointed
out by Gannon.

At the next level of the hierarchy we have the $sum$ of the next Mersenne
prime: $ 2^{127} - 1 $ to the number $ 137$ obtained by the previous
sums of the Mersenne primes $ 3 + 7 + 127 = 137 $. Hence this sum allows
the bit-string physics evaluation of the Planck mass to proton mass ratio
in terms of the Newton constant $ G_N $:

$$ ({ M_{Planck} \over m_{proton}})^2 = {\hbar c \over G_N m_p^2} =
[(2^{127} - 1) + (137)] [1 - {1 \over 3 \times 7 \times 10}] =
1.6933 \times 10^{38}. \eqno (65) $$
compared to the experimental value of $ 1.69358 \times 10^{38} $.
The Fermi constant; the Electro-weak mixing angle; the proton-electron
mass ratio; the pion-electron mass ratio; the muon-electron mass ratio;
the pion-nucleon coupling constant; the dark-matter to baryon
ratio....have all been calculated to remarkable precision. Is this just
numerology? Or there is some deep physical organizing principle in
Nature that can be described by these fractal strings? If nature chose
the Golden mean to scale its fractal strings why did it do so? The
Golden Mean has been known to ancient Greeks, to artists of the
Renaissance,....Perhaps one must invoke a deeper principle of harmony
operating in Nature that selected the Golden Mean over all other numbers.
The Golden Mean obeys striking self-similar decompositions involving the
unit number $ 1 $. Wheeler speculated in the past that information theory
may lie at the heart of things. Perhaps Nature's code involves the Golden
Mean at its core.

It has been know for some time that there are very strong resemblances
between Hadronic Physics and Set theory based on the Banach-Tarski
paradox which is one of the most shocking results in Mathematics. It says
that a solid ball in $R^3$ can be broken into $5$ pieces that can be
re-arranged using rigid motions of $3$-space to form $two$ balls each of
which is the $same$ size as the original...These pieces are
non-measurable sets and their construction requires the use of the Axiom
of Choice.

Augenstein has given convincing arguments [20] that every observed
strong interaction involving a hadron-reaction can be envisaged as a
paradoxical decomposition or a sequence of paradoxical decompositions.
The role of non-Abelian groups in both hadronic physics and this
paradoxical decomposition is one mathematical link which connects these
two areas.
One can evision the nucleon/baryon composed of $3$ quarks and mesons
compose of $2$ quarks giving a total of $5$ pieces involved in the
Banach-Tarski paradoxical decomposition. Mycielski and Wagon have
developed a computer program [21] that allows one to see the essence of
this paradox using only triangles in the hyperbolic plane $ H^2$. They
have shown how tilings of the $H^2$ exist where starting from $3$
congruent regions of the $H^2$, say Red, Blue, Green (three quarks) one
can, by a mere change of the viewpoint of the tiling, to make it clear
that each set is congruent also to its complement ! Thus, these sets are
simultaneously $ 1/2 $ and $ 1/3$ of the hyperbolic plane !
One can then evision the three quarks inside the baryon broken up into
myriads of different pieces into three distinct regions of Red, Blue and
Green. This would represent a hierarchical structure of the quarks
themselves. Klein's celebrated tessellation of the $ H^2$ involved using
the (hyperbolic) triangle whose angles were $ (\pi/2, \pi/3,\pi/ 7)$.
The fact that the primes $ (2, 3, 7)$ appear is also very
suggestive. Are there other tessellations of the hyperbolic plane
involving Mersenne primes?

A hierarchy of scales using Mersenne primes has also been pointed out by
Pitkanen [17]. He has a hierarchy given by:
$$ L = {\sqrt M_p} L_o. ~~~ L_o = 137 \times 10^2~ Planck~scale. \eqno (
66) $$
The Merseene primes associated with:
$$ p = 127, 107, 89, 61, 2. \eqno (67) $$
represent the Compton wavelengths of the electron, proton, $W$ boson and
the fluctuon particle [3] and the GUT scale respectively.

To finalize we will mention how Mersenne primes are
connected to a perfect number [19]. A perfect number $ P $
is that number that can be written as the sum of $all$ of
its positive divisors. For example:
$$ 6 = 2 \times 3 = 1 + 2 + 3. ~~~ 28 = 4 \times 7 = 1+2 + 4 + 7 + 14
..... \eqno (68) $$
When I speak of a perfect number it is clear that I refer
to even perfect numbers. It is unknown if there exist odd
perfect numbers. A perfect number can be written in terms
of a Mersenne prime as:
$$ P (p) = (2^{p -1}) (2^p - 1) = 2^{p -1} M_p. \eqno (69) $$

The curious fact is that the perfect numbers $ P(p) $ coincides with the
number of generators of the groups $ SO(2^p) $ given by

$$ {1 \over 2} (2^p) (2^p - 1) = P(p). \eqno (70) $$

The most relevant example is the number of generators of $ SO(32) =
SO(2^5) $ which is the group associated with the anomaly free open
superstring in $ D = 10 $ and has $ 496$ generators:

$$ {1\over 2} 32 \times 31 = 16 \times 31 = 496 ~ generators. ~~~ 31 =
2^5 - 1 = Mersenne~ prime. $$

It is undoubtedly that number theory will bring us more surprises in the
future.

\bigskip

\centerline {\bf Final Comments}

\bigskip

We are going to add a few important remarks pertaining some
of the topics discussed in this work. The first remark
concerns the nature of (the inverse of) fine structure
constant $\alpha_0 \approx 137$. It is well known that the
coupling constants are not constant, they are running, due
to quantum effects imposed by the Renormalization Group
program. Therefore from this point of view there is nothing
$fundamental$ to the number $\alpha_0$ since it will
$change$ with scale. The fundamental nature of $\alpha_0$
was noticed by Dyson [15], who interpreted it as a true
$counting$ number, rather than a pure dimensional
interpretation of the type given by Eddington-El Naschie.
Dyson noticed that if one had $137$ electron-positron
pairs, the perturbation series of QED will cease to be
uniformly convergent. Hence one should see $\alpha_0$ as a
$counting$ number. This is the basis of the bit-string
physics combinatorial hierarchy aproach to the fundamental
constants in Nature based on the doubly Mersenne prime
hierarchy.

Another place where the prime number $137$ appears as a
$counting$ number is algebraic geometry. Saniga [22]
studied the sequence of integers generated by the $number$
of conjugated pairs of homaloidal nets of plane algebraic
curves of even order and, remarkably, found it to provide
an exact integer match to El-Naschie's hierarchy of
dimensions inspired from transfinite heterotic strings, $
10 \times 2 (1 + \phi)^n $, $up to$ the El
Naschie-Selvam-Fadnavis value: $ 20 (1 + \phi)^4 = 100 +
60 \phi = 137.082...$. This indicates that also some sort of
algebraic geometrical constraint exists in Physics which
selects the number $ 137$. For more details concerning the
arithmetic of plane Cremona transformations and its
relation to the transfinite string hierarchy see [22].

So far we have been speaking of the hierarchy of dimensions
induced by the core dimension $ D = 10 $ of the
superstrings. The transfinite $M$ theory [18] relies on
the dimensions of the alleged anomaly-free supermembrane: $
D = 11 $ and has for hierarchy the sequence $ 11 \times 2
(1 + \phi)^n$. The latter was consistent with the
$p$-adic-like expansion of the inverse fine structure
constant given by eq-(56). The hierarchy of dimensions of
transfinite $M$ theory is consistent with the numbers of
heptagons appearing in the heptagonal hyperbolic honeycomb
tessallation associated with Klein's quartic curve. The
number of heptagons grows as $7$ times the Fibonacci
sequence [23]:

$$ 7 \times ~ Fibonacci = 7, 7, 14, 21, 35, 56, 91, 147, 238,
385, 623... $$
this sequence of numbers is close in its integer-valued
part to the transfinite $M$ theory hierarchy for the
following values of $ n = 4, 3, 2, 1, 0, -1, -2 $
respectively:
$$ 11 \times 2 (1 +\phi)^n \Rightarrow, 150.8; 93.2; 57.6;
35.6; 22;13.6; 8.4 $$
For the crucial role of Borcherds symmetries in $M$ theory,
Del Pezzo surfaces and the enhanced hierarchy of dimensions
induced by $ 11 \times (1 + \phi)^n$ see [24, 28].

Our second topic consists now on the role of even perfect
numbers in string theory. We have discussed that the number
of generators of the algebra $ SO(2^5) = SO(32)$ was given
by an even perfect number since $ 31 $ was a Mersenne prime
$ 2^5 - 1 $. The number of generators was $ 496$ which
coincides with the dimension of $ E_8 \times E_8 = 248 + 248$.
The $ SO(32) $ and $ E_8 \times E_8$ were the anomaly free
groups in superstring theory.

Another important group related to the $unique$
tadpole-free bosonic string theory is the $ SO(2^{13}) =
SO(8192)$ bosonic string theory compactified on the $ E_8
\times SO(16) $ lattice which determines the anomaly free
Chan-Paton group of the type $ I$ string theory, up to
one-loop. It is the bosonic ancestor of closed and open
fermionic strings [25]. The number of generators of $ SO(
2^{13})$ is once again given by an (even) perfect number
since the number $ 2^{13} - 1 $ is a also Mersenne prime.
I have been recently informed by Saniga [26] that the
expression for the Mersenne primes, $ 2^p -1 $, is
identical with that for the number of points in the
$p-1$-dimensional projective space over the simplest of
Galois fields, viz. the field featuring just two elements,
and that the sequence of the $four$ doubly Mersenne primes
could bear a connection to the $four$ families of symmetric
homaloidal nets. Another interesting question raised by
[26] is if there are relations between the Mersenne prime
hierarchy and the prime-producing quadratic polynomials
[27].

Another application of the even perfect numbers is in the
theory of the Riemann's theta functions with
characteristics. They appear naturally in the superstring
perturbation theory defined over super-Riemann surfaces of
arbitrary genera. The even/odd characteristics are
associated with the even/odd spin structures of the
super-Riemann surface. The number of odd characteristics in
terms of the genus number $ g$ is given by $(2^{g -1})(2^g
- 1)$ which is once again the expression for an even
perfect number if $ (2^g - 1) $ is prime. For example, for
Klein's quartic curve that has genus $ 3 $ this formula
yields $ 4\times 7 = 28 $. It has been shown by Weber that
the number of odd characteristics of the Riemann's theta
with characteristics is intimately linked to the $28$
bitangents to the Klein's quartic curve.

These $28$ bitangents are connected to the $27$ lines on a
cubic surface [29] and their automorphism group is
isomorphic to the Weyl group of the exceptional Lie algebra
$ E_7 $. It is well known that the root spaces of the
$E_n$ Lie groups appear in the second integral cohomology
of regular, complex, compact, Del Pezzo surfaces. The full
Borcherds' root lattices are exactly the second integral
cohomology of Del Pezzo surfaces [24]. Another interesting
application of the algebra of the configurations of the
lines on Del Pezzo surfaces in physics is outlined in [22].

To finalize we point out the need to construct an
Arithmetic Quantum Field theory based on the fractal string
construction [1] and Arithmetic groups. The fractal
string/spray geometric counting function is the arithmetic
analog of the partition function in Statistical Mechanics
and the $discrete$ version of the string/brane actions!
This is why one could construct an Arithmetic QFT based on
such geometric counting function! For example, one may
relate the Tsallis Entropy written in terms of a sequence
of probabilities $ p_i $ as:

$$ {\cal H} (s) \equiv = {1 - \sum (p_i)^s \over 1 - s}. ~~~
{\cal H}(s = 1) = \sum (p_i) (ln ~ p_i) = Shannon. ~~~ \sum
p_i = 1. $$
The expression for the Tsallis entropy can be extended to
the fractal string/spray geometric counting function $ Z(s)
$ in eq-(1), by imposing the correspondence $ p_i
\leftrightarrow l_i / \sum l_i $ and writing:

$$ {\cal H}_{fractal} = {1 - \sum (l_i)^s \over 1 - s} =
{1 - Z(s) \over 1 - s} $$
The expectation value of the logarithm of the fractal
string length would be:

$$ < ln ~ l > = {\sum (l_i) (ln ~ l_i) \over \sum l_i}
\sim {\cal 
H}_{string} (s = 1) = {\partial Z(s) \over \partial s}|_{s
= 1}. $$

Since ordinary QFT is based on Lie group symmetries we
expect that Arithmetic QFT will be based on Arithmetic
groups. For the role of Hurwitz groups in surfaces, like
that corresponding to Klein's quartic, see [23]. Klein's
quartic is associated with the first of the Hurwitz groups.
For this reason it is called also a Hurwitz-Shimura modular
curve.

An impending question is to analyze the mass spectrum
associated with these fractal strings/sprays. Since the
dimension of spacetime is connected with the energy
content, and the latter with mass, it is very reasonable
to assume that the spectrum of masses is correlated with
the spectrum of dimensions. In particular, with the
$imaginary$ components of the dimension; i.e. with the
$oscillations$ in the geometry of fractal strings/sprays.
Hence, the towers of masses should correspond to the
towers of the imaginary components of the dimension, since
the latter encode the oscillations in the geometry of the
fractal strings. Hence the formula for the mass spectrum
associated with a family of fractal strings/sprays,
parametrized by the prime $ p$ and the dimension of the
sets $ {\cal E}^{(j)} = (1 +\phi)^{j -1} $ is given by
eq-(47):

$$ m_o {2 \pi n \over ln~p} (1 + \phi)^{j - 1} = M_n(
p, j). ~~~ n = 
0, 1, 2, 3.... ~~~ p = prime. $$
where $m_o$ is the fiducial reference mass (say the
electron mass). It remains to see if this ansatz for the
masses is compatible with recent results obtained by El
Naschie by choosing the values of $ p, n $ appropriately.

\bigskip

\centerline {\bf Acknowledgments}
\bigskip

I am indebted to Laurent Nottale for discussions on the complex dimensions
and logarithmic periodic laws. To Alex Granik, Jorge Mahecha and Metod
Saniga for their help in correcting and preparing this manuscript. To L.
Perelman, S. Perelman and M. Bowers for their hospitality.

\vfill\eject

\centerline {\bf References}

\parindent=0pt
\bigskip

1- M. Lapidus, M. van Frankenhuysen: ``Fractal Geometry and Number Theory,
Fractal strings and zeros of the zeta function''. Birkhauser 2000.

2- M. S. El Naschie: Chaos, Solitons and Fractals, {\bf 9} (3) (1998)
517.

3- M. Conrad: Chaos, Solitons and Fractals, {\bf 7} (5) (1996) 725.

4- L. Nottale: ``La Relativite dans tous ses Etats''. Hachette Literature
Paris 1999.

5- L. Nottale, J. Chaline, P. Grou: ``Les arbres de l'Evolution''
Hachette Literature Paris (2000).

6- D. Sornette: Physics Reports {\bf 297} (1998) 239-270.

7- M. Nauenberg: J. Phys. A, Math, General {\bf 8} (1975) 925.

G. Jona-Lasinio: Il Nuovo Cimento {\bf 26 B} (1975) 99.

8- J. von Neumann: ``Continuous Geometry'' Princeton University Press,
second edition 1988.

9- M. S. El Naschie: Chaos, Solitons and Fractals {\bf 13} (2002) 1935.

V. S. Krushkal: Geometry and Topology {\bf 4} (2000) 397.

10- M. Baake, R. Moody, M. Schlottman, math-ph/9901008 and
math-ph/9809008.

M. Baake, R. Moody: ``Similarity subgroups and semigroups in
Quasi-crystals'' eds J. Patera. AMS, Rhode Island 1990.

11- V. Vladimorov, I. Volovich, I. Zelenov: ``p-adic numbers in Math.
Physics'' World Scientific Singapore 1994.

L. Brekke, P. Freund: Phys. Reports {\bf 231} (1993) 1.

12- C. Castro, A. Granik: ``Planck scale relativity and variable fine
structure from C-spaces'' to appear.

13- M. Selvam, S.Fadnavis: Chaos, Solitons and Fractals {\bf 10} (8)
(1999) 1321.

14- M. S. El Naschie: Chaos, Solitons and Fractals {\bf 12} (2001) 969.

15- P. Noyes: ``Bit strings Physics'' hep-th/9707020.

P. Noyes: ``Bit string physics,a discrete and finite approach to natural
philosophy''. Series of Knots in Physics {\bf 27}, World
Scientific Singapore 2001. Eds by J. C van der Berg.

16- T. Smith: ``From Sets to Quarks'' hep-ph/9708379.

17- M. Pitkannen: Topological Geometrodynamics I and II:

Chaos, Solitons and Fractals {\bf 13} (6) (2002) 1205 and 1217.

18- C. Castro: Chaos, Solitons and Fractals {\bf 13} (2002) 203.

19- Caldwell, web page on primes: http://www.utm.edu/primes.

20- B. Augenstein: Int. Jour. of Theor. Physics {\bf 23} (12) (1984)
1197.

21- S. Wagon: ``The Banach Tarski paradox'' Cambridge University Press
and Chapter 21 of the textbook of Mathematica.

22- M. Saniga: ``Arithmetic of plane Cremona transformations
and the dimensions of transfinite heterotic string
space-time'', Chaos, Solitons and Fractals 2002;13:1537-40.

 M. Saniga: ``A further note on a formal relationship
between the arithmetic of homaloidal nets and the
dimensions of transfinite space-time'', Chaos, Solitons and
Fractals 2002;13:1571-3.

23- S. Levy (ed.): ``The Eightfold Way: The Beauty of
Klein's Quartic Curve'', Mathematical Sciences Research
Institute Publications. Cambridge: Cambridge University
Press; 1999.

24- P. Henry-Labordere, B. Julia and L. Paulot: ``Borcherds
symmetries in M theory'', hep-th/0203070.

25- F. Englert, L. Houart and A. Taormina: ``The Bosonic
sector of closed and open fermionic strings'',
hep-th/0203098.

26- M. Saniga: Private Communication.

27- R. A. Mollin: ``Quadratics'' New York: CRC Press; 1995.

28- M. Saniga: ``Lines on Del Pezzo surfaces and transfinite
heterotic string space-time'', Chaos, Solitons and Fractals
2002;13:1371-3.

29- A. Henderson: ``The twenty-seven lines upon a cubic
surface''. Cambridge: Cambridge University Press; 1911.

\bye